\documentclass[10pt]{amsart}

\usepackage{hyperref}

\usepackage{amssymb}
\usepackage{amsmath}
\usepackage{graphicx}
\usepackage{tikz}
\usepackage{amsthm}

\usepackage{bm}

\let\newpf\proof \let\proof\relax

\newcommand{\bt}{\begin{thm}}
\newcommand{\et}{\end{thm}}

\newcommand{\bl}{\begin{lemma}}
\newcommand{\el}{\end{lemma}}

\newcommand{\beq}{\begin{eqnarray}}
\newcommand{\eeq}{\end{eqnarray}}

\def\be{\begin{equation}}
\def\ee{\end{equation}}

\def\ba{{\begin{align}}}
\def\ea{{\end{align}}}

\def\0{{\mathbf 0}}

\newtheorem{thm}{Theorem}[section]

\newtheorem{lem}[thm]{Lemma}
\newtheorem{lemma}[thm]{Lemma}

\theoremstyle{remark}

\numberwithin{equation}{section}

\def\note#1
{\marginpar

{\tiny $\leftarrow$
\par
\hfuzz=20pt \hbadness=9000 \hyphenpenalty=-100 \exhyphenpenalty=-100
\pretolerance=-1 \tolerance=9999 \doublehyphendemerits=-100000
\finalhyphendemerits=-100000 \baselineskip=6pt
#1}\hfuzz=1pt}

\renewcommand{\mod}{\operatorname{mod}}

\newcommand{\R}{{\mathbb R}}
\newcommand{\T}{{\mathbb T}}

\newcommand{\Z}{{\mathbb Z}}

\def\B0{{\bold{0}}}

\newcommand{\tthe}{\tilde{\theta}}
\newcommand{\bthe}{\bm{\theta}}
\newcommand{\tbt}{\tilde{\bm{\theta}}}

\newcommand{\tbb}{\tilde{\bm{\beta}}}

\catcode`\@=12

\def\Empty{}
\newcommand\oplabel[1]{
  \def\OpArg{#1} \ifx \OpArg\Empty {} \else
  	\label{#1}
  \fi}

\newcommand{\comm}[1]{}
\newcommand{\comment}[1]{}

\begin{document}

\title[]{Discrete Bethe-Sommerfeld conjecture}

\author{Rui Han and Svetlana Jitomirskaya}


\begin{abstract}
In this paper, we prove a discrete version of the Bethe-Sommerfeld conjecture. 
Namely, we show that the spectra of multi-dimensional discrete periodic Schr\"odinger operators on $\Z^d$ lattice with sufficiently small potentials contain at most two intervals. 
Moreover, the spectrum is a single interval, provided one of the
periods is odd, and can have a gap whenever all periods are even. 
\end{abstract}

\maketitle

\section{Introduction}
Bethe-Sommerfeld conjecture states that for $d\geq 2$ and any periodic function $V:\R^d\rightarrow \R$, the spectrum of the continuous Schr\"odinger operator:
\begin{align*}
-\Delta+V
\end{align*}
contains only finitely many gaps, so no gaps for large energies. This conjecture has been studied extensively with many important advances \cite{DT, HM, Kar, PSDuke, PSAHP, PS, S79, S84, S85}.
Finally, Parnovski \cite{P}, proved it in any dimension $d\geq 2$, under smoothness conditions on the potential $V$ (see \cite{V} for an alternative approach).

In this paper, we consider a discrete version of this conjecture.
A discrete multi-dimensional periodic Schr\"odinger operator on $l^2(\Z^d)$ is given by:
\begin{align}\label{defH}
(H_{V}u)({\bm{n}})=\sum_{|\bm{m}-\bm{n}|=1}u({\bm{m}})+ V(\bm{n})u({\bm{n}}),
\end{align}
where $|\bm{m}-\bm{n}|=\sum_{i=1}^d |m_i-n_i|$. We assume $V(\cdot)$ is a bounded real-valued periodic function on $\Z^d$ with period $\bm{q}=(q_1, q_2,..., q_d)$, namely, 
$V(\bm{n}+q_i\bm{b}_i)=V(\bm{n})$, with $\{\bm{b}_i\}_{i=1}^d$ being the standard basis for $\R^d.$ 
\footnote{The most general periodic case may seem to be
  $V(\bm{n}+\bm{w}_i)=V(\bm{n})$, where $\bm{w}_i\in \Z^d$,
  $i=1,...,d$, are linearly independent vectors. This however reduces
  toour assumption because such operators are periodic with
  period $\bm{q}=(\det{W},..., \det{W})$, where $W$ is the matrix with
  $\bm{w}_i$ as columns.}.In the high energy regime continuous
Schr\"odinger operators can be viewed as perturbations of the free
Laplacian. In this sense the proper discrete analogy of the
Bethe-Sommerfeld conjecture is absence of gaps for small coupling
discrete periodic operators.

The discrete Bethe-Sommerfeld conjecture has been proved for $d=2$ by
Embree-Fillman \cite{EF}, with a partial result (for coprime periods)
earlier by Kr\"uger \cite{Kru}. The approach of \cite{EF} runs into
combinatorial/algebraic difficulties for $d>2.$
Here we prove this conjecture for arbitrary dimensions:
\begin{thm}\label{main}
Let $d\geq 2$ and a period $\bm{q}=(q_1, q_2, ..., q_d)$ be given. There exists a constant $c_{\bm{q}}>0$ such that the following statements hold:
\begin{enumerate}
\item If $\|V\|_{\infty}\leq c_{\bm{q}}$, then the spectrum of $H_V$ contains at most two intervals.
\item If at least one of $q_i$ is odd, and $\|V\|_{\infty}\leq c_{\bm{q}}$, then the spectrum of $H_V$ is a single interval.
\end{enumerate}
\end{thm}

Our result is sharp in the sense that if all the $q_i$'s are even, then there exists $V$ (see example in Section \ref{countersec}) with {\it minimal period} $\bm{q}$, and arbitrarily small $\|V\|_{\infty}$ such that $\Sigma(H_{V})$ contains {\it exactly} two intervals. 
The example we give is a modification of Kr\"uger's example \cite{Kru}, in which $V(\bm{n})=\delta (-1)^{|\bm{n}|}$ has minimal period $(2,2,...,2)$.
Also it is well-known that both $d\geq 2$ and the smallness of $\|V\|_{\infty}$ are needed.

The strategy of our proof relies on analysing the overlaps of adjacent bands of the spectrum. 
We refer the readers to \cite{Kru} for detailed background on discrete multi-dimensional Schr\"odinger operators.  
Here we only introduce some notations and known results.
Let us denote the spectrum of $H$ by $\Sigma(H)$.
By Floquet-Bloch decomposition, $\Sigma(H_V)$ can be decomposed into $\cup_{\bm{\theta}\in \Theta}\Sigma(H_V^{\bm{\theta}})$, where $\Theta=\{\bm{\theta}=(\theta_1, \theta_2, ..., \theta_d): 0\leq \theta_i< \frac{1}{q_i},\ 1\leq i\leq d\}$ is a $d$-dimensional torus (by gluing $0$ and $\frac{1}{q_i}$ together in the $\bm{b}_i$ direction),
and $H_{V}^{\bm{\theta}}$ is the periodic Schr\"odinger operator with the following boundary condition:
\begin{align*}
u_{\bm{n}+q_i\bm{b}_i}=e^{2\pi i q_i\theta_i} u_{\bm{n}}.
\end{align*}

Each operator $H_V^{\bm{\theta}}$ clearly has $Q=\prod_{i=1}^d q_i$ eigenvalues, which we will order in the {\it decreasing} order and denote them by 
$E_V^1(\bthe)\geq E_V^2(\bthe)\geq \cdots \geq E_V^Q(\bthe)$. 
Let $F_V^k=\cup_{\bthe\in \Theta}E_V^k(\bthe)$ be the $k$-th band of the spectrum.
Theorem \ref{main} is thus reduced to proving non-empty overlaps of arbitrary two adjacent bands, with only possible exception around the point $0$.
Employing a standard perturbation argument (see Theorem \ref{perturb}), this is made possible via proving non-empty overlaps of the {\it interiors} of adjacent bands of the free Laplacian $H_0$. 
Two of our key lemmas are as follows:
\begin{lemma}\label{Enot0}
If $E\in (-2d, 2d)\setminus \{0\}$, then $E\in \mathrm{int}(F^k_0)$ for some $1\leq k\leq Q$.
\end{lemma}

\begin{lemma}\label{E=0}
If at least one of $q_i$'s is odd, then $0\in \mathrm{int}(F^k_0)$ for some $1\leq k\leq Q$.
\end{lemma}
We will prove Lemma \ref{Enot0} in Section \ref{Enot0sec} and Lemma \ref{E=0} in Section \ref{Eequal0sec}. 
Different from the existing $d=2$  proofs in \cite{Kru, EF}, our argument proceeds by contradiction. 
Namely we assume $E^{k_0}_0(\tthe)=\min F^{k_0}_0=\max F^{k_0+1}_0$ for certain $k_0$, and then apply a novel {\it perturb-and-count} technique.
We perturb the phase $\tthe$ and count the number of eigenvalues that move up and down. 
It is then argued that different chosen directions lead to different numbers of eigenvalues that go up/down, hence a contradiction.

\section{Preliminaries}
For $\bthe, \tbt\in \Theta$, let $\|\bthe-\tbt\|_{\Theta}$ be the torus distance between them, defined by
\begin{align*}
\|\bthe-\tbt\|^2_{\Theta}=\sum_{i=1}^d \|\theta_i-\tthe_i\|_{\T_i}^2,
\end{align*}
where $\|\theta\|_{\T_i}:=\mathrm{dist}(\theta, \frac{1}{q_i}\Z)$.

\subsection{Spectrum of the free Laplacian}
It is a well-known result that the spectrum of the free Laplacian $H_0$ is a whole interval:
\begin{align}
\Sigma(H_0)=[-2d, 2d].
\end{align}

By Floquet-Bloch decomposition,
\begin{align}
\Sigma(H_0)=[-2d, 2d]=\cup_{\bm{\theta}\in \Theta}\Sigma(H_0^{\bm{\theta}}).
\end{align}
Furthermore, each $\Sigma(H_0^{\bm{\theta}})$ can be written down explicitly,
\begin{align}\label{spectrumlambda=0}
\Sigma(H_{0}^{\bm{\theta}})=\left\lbrace e^{\bm{l}}_0(\bm{\theta}):=2\sum_{i=1}^d \cos{2\pi (\theta_i+\frac{l_i}{q_i})}\right\rbrace_{\bm{l}\in \Lambda},
\end{align}
where $\Lambda=\{\bm{l}=(l_1, l_2, ..., l_d): 0\leq l_i\leq q_i-1,\ 1\leq i\leq d\}$.


\section{Proof of Theorem \ref{main}}
We say the bands $\{F_k\}_{k=1}^Q$ of $H$ are $\delta$-overlapping if $\max F^{k+1}-\min F^k\geq \delta$ for any $1\leq k\leq Q-1$.
Theorem \ref{main} follows from a quick combination of Lemmas \ref{Enot0}, \ref{E=0} with Hausdorff continuity of the spectrum. 
The form of continuity convenient to us is presented in:
\begin{thm}\label{perturb}(\cite{Kru}, Theorem 3.8)
Let the bands of $H$ be $\delta$-overlapping. Then the bands of $H+V$ are
$\delta-2\|V\|_{\infty}$-overlapping.
\end{thm}
$\hfill{} \Box$

\section{Proof of Lemma \ref{Enot0}}\label{Enot0sec}
Our strategy is to prove by contradiction, namely we assume $\min F^{k_0}_0 = \max F^{k_0+1}_0\neq 0$ for some $1\leq k_0\leq Q$ and try to get a  contradiction. 
Without loss of generality, we assume $\min F^{k_0}_0 = \max F^{k_0+1}_0>0$.


We will use the following elementary lemma, whose proof will be included in the appendix.
\begin{lem}\label{thereexiststheta}
Let $d\geq 2$.
For any $E\in (-2d, 2d)$, there exists $\bthe=(\theta_1, \theta_2,..., \theta_d)$ with $\theta_i\in [0, 1)$ such that
\begin{align*}
\left\lbrace
\begin{matrix}
\sum_{i=1}^d 2\cos{2\pi \theta_i}=E,\\
\\
\sum_{i=1}^d \sin{2\pi \theta_i}=0,\\
\\
\sum_{i=1}^d \sin^2 {2\pi \theta_i}\neq 0.
\end{matrix}
\right.
\end{align*}
\end{lem}


Now let us prove Lemma \ref{Enot0}.

First, by Lemma \ref{thereexiststheta}, there exists $\tbt=(\tthe_1, \tthe_2, ..., \tthe_d)\in \Theta$ and $\bm{l}^{(1)}=(l^{(1)}_1, l^{(1)}_2, ..., l^{(1)}_d)\in \Lambda$ such that 
\begin{align}\label{l1orthoto}
\left\lbrace 
\begin{matrix}
\min F^{k_0}_0=\sum_{i=1}^d 2\cos{2\pi (\tthe_i+\frac{l^{(1)}}{q_i})}=e^{\bm{l}^{(1)}}_0(\tbt),\\
\\
0=\sum_{i=1}^d \sin {2\pi (\tthe_i+\frac{l^{(1)}}{q_i})},\\
\\
0\neq \sum_{i=1}^d \sin^2 {2\pi (\tthe_i+\frac{l^{(1)}}{q_i})}.
\end{matrix}
\right.
\end{align}


Next, let us choose $\bm{l}^{(2)}, \bm{l}^{(3)},..., \bm{l}^{(r)} \in \Lambda$ (if any) be {\it all} the vectors in $\Lambda$ such that
\begin{align*}
e^{\bm{l}^{(1)}}_0(\tbt)=e^{\bm{l}^{(2)}}_0(\tbt)=\cdots=e^{\bm{l}^{(r)}}_0(\tbt).
\end{align*}
Then clearly they are $E_0^{k_0-s}(\tbt)=\cdots=E_0^{k_0}(\tbt)=\cdots =E_0^{k_0+r-s-1}(\tbt)$, for some $0\leq s\leq r-1$. 
And also we have $E_0^{k_0-s-1}(\tbt)>E_0^{k_0-s}(\tbt)$, $E_0^{k_0+r-s-1}(\tbt)>E_0^{k_0+r-s}(\tbt)$. 
By the continuity of each eigenvalue, we could choose $\epsilon>0$ small enough, such that for any $\|\bthe-\tbt\|_{\Theta}<\epsilon$, we always have 
\begin{align}
E_0^{k_0-s-1}(\bthe)>E_0^{k_0-s}(\bthe)\ \mathrm{and}\ E_0^{k_0+r-s-1}(\bthe)>E_0^{k_0+r-s}(\bthe).
\end{align}

Let $J_0\geq 0$ be the number of $j$'s such that $\nabla e^{\bm{l}^{(j)}}_0(\tbt)=\bm{0}$.
For $\bm{\beta}\in \R^d$, we also introduce $J_{\bm{\beta}}$ and $J_{\bm{\beta}}^0$:
let $J_{\bm{\beta}}$ be the number of $j$'s such that $\bm{\beta}\cdot \nabla e^{\bm{l}^{(j)}}_0(\tbt)>0$, and $J_{\bm{\beta}}^0$ be the number of $j$'s such that $\nabla e^{\bm{l}^{(j)}}_0(\tbt)\neq \bm{0}$ and $\bm{\beta}\cdot \nabla e^{\bm{l}^{(j)}}_0(\tbt)=0$.\\
Perturbing $e_0^{\bm{l}^{(j)}}(\tbt)$ along the direction of $\bm{\beta}$ we get:
\begin{align}
e^{\bm{l}^{(j)}}_0(\tbt+t\bm{\beta})
=&e^{\bm{l}^{(j)}}_0(\tbt)+t \bm{\beta}\cdot \nabla e^{\bm{l}^{(j)}}_0(\tbt)+O(t^2)\label{taylort} \\
=&e^{\bm{l}^{(j)}}_0(\tbt)+ t\bm{\beta}\cdot \nabla e^{\bm{l}^{(j)}}_0(\tbt)+ \frac{t^2}{2} \left(-4\pi^2 \sum_{i=1}^d 2\cos{2\pi (\tthe_i+\frac{l^{(j)}_i}{q_i})} {\beta}_i^2 \right)+O(t^3)\label{taylor21}.
\end{align}
\

\subparagraph{Step 1}\

Let $\tbb=\frac{1}{\sqrt{d}}(1,1,...,1)$. By (\ref{l1orthoto}), we have 
\begin{align}\label{l1orthto2nd}
\tbb\cdot \nabla e^{\bm{l}^{(1)}}_0(\tbt)=0\ \ \mathrm{and}\ \ \nabla e^{\bm{l}^{(1)}}_0(\tbt)\neq \bm{0},
\end{align}
which implies $J_{\tbb}^0\geq 1$.

By (\ref{taylor21}) for $j$ such that $\tbb\cdot \nabla e^{\bm{l}^{(j)}}_0(\tbt)=0$ (in total $J_0+J^0_{\tbb}$ many such $j$'s), we have
\begin{align}
e^{\bm{l}^{(j)}}_0(\tbt+t\tbb)
=(1-\frac{2\pi^2}{d} t^2)e^{\bm{l}^{(j)}}_0(\tbt)+O(t^3)
<e^{\bm{l}^{(j)}}_0(\tbt)\label{taylor2},
\end{align}
for $|t|$ small enough. Let us mention that in (\ref{taylor2}), we used the fact that $e^{\bm{l}^{(j)}}_0(\tbt)=\min F^{k_0}_0> 0$.


Now combine (\ref{taylort}) with (\ref{taylor2}). On one hand, we have, for $\epsilon>t>0$ small enough,
\begin{itemize}
\item there are $J_{\tbb}$ many $j$'s such that $E^{k_0-s-1}(\tbt+t\tbb)>e^{\bm{l}^{(j)}}_0(\tbt+t\tbb)>e^{\bm{l}^{(j)}}_0(\tbt)=\max F^{k_0+1}_0$, 
thus $J_{\tbb}\leq (k_0+1)-(k_0-s-1)-1=s+1$.
\item for the other $r-J_{\tbb}$ many $j$'s, we have $E^{k_0+r-s}(\tbt+t\tbb)<e^{\bm{l}^{(j)}}_0(\tbt+t\tbb)<e^{\bm{l}^{(j)}}_0(\tbt)=\min F^{k_0}_0$,
so $r-J_{\tbb}\leq (k_0+r-s-1)-(k_0+1)+1=r-s-1$.
\end{itemize}
Thus 
\begin{align}\label{relation11}
J_{\tbb}=s+1.
\end{align}

On the other hand, for $0>t>-\epsilon$ small enough, we have,
\begin{itemize}
\item there are $r-J_{\tbb}-J^0_{\tbb}-J_0$ many $j$'s such that $E^{k_0-s-1}(\tbt+t\tbb)>e^{\bm{l}^{(j)}}_0(\tbt+t\tbb)>e^{\bm{l}^{(j)}}_0(\tbt)=\max F^{k_0+1}_0$,
\item for the other $J_{\tbb}+J^0_{\tbb}+J_0$ many $j$'s, we have $E^{k_0+r-s}(\tbt+t\tbb)<e^{\bm{l}^{(j)}}_0(\tbt+t\tbb)<e^{\bm{l}^{(j)}}_0(\tbt)=\min F^{k_0}_0$.
\end{itemize}
Thus 
\begin{align}\label{relation12}
J_{\tbb}+J^0_{\tbb}+J_0=r-s-1.
\end{align}
Combining this with (\ref{relation11}), we have, 
\begin{align}\label{relation13}
r-2s=J^0_{\tbb}+J_0+2.
\end{align}


\

\subparagraph{Step 2}\ 

We choose $\bm{\beta}\in \R^d$, $\|\bm{\beta}\|_{\R^d}=1$, such that $\bm{\beta}\cdot \nabla e^{\bm{l}^{(j)}}_0(\tbt)\neq 0$ for any $1\leq j\leq r$ with 
$\nabla e^{\bm{l}^{(j)}}_0(\tbt)\neq \bm{0}$, and satisfies the following:
\begin{align}\label{choosebeta}
\sum_{i=1}^d 2|\beta_i^2-\frac{1}{d}|<\frac{1}{2d}\min F^{k_0}_0.
\end{align}
Inequality (\ref{choosebeta}) basically says $\bm{\beta}\sim \tbb$.

For $j$ such that $\nabla e^{\bm{l}^{(j)}}_0(\tbt)=\bm{0}$, we have, by (\ref{taylor21}),(\ref{choosebeta})
\begin{align}
e^{\bm{l}^{(j)}}_0(\tbt+t\bm{\beta})
=&e^{\bm{l}^{(j)}}_0(\tbt)+\frac{t^2}{2} \left(-\frac{4\pi^2}{d} e^{\bm{l}^{(j)}}_0(\tbt)+4\pi^2\sum_{i=1}^d 2\cos{2\pi (\tthe_i+\frac{l^{(j)}_i}{q_i})}(\frac{1}{d}-\beta_i^2)\right)+O(t^3)\notag\\
\leq &(1-\frac{\pi^2}{d} t^2)e^{\bm{l}^{(j)}}_0(\tbt)+O(t^3)\label{taylor31}\\
<&e^{\bm{l}^{(j)}}_0(\tbt).\label{taylor3}
\end{align}

Combining (\ref{taylort}) with (\ref{taylor3}), on one hand, we have that for $\epsilon>t>0$ small enough,
\begin{itemize}
\item there are $J_{\bm{\beta}}$ many $j$'s such that $E^{k_0-s-1}(\tbt+t\bm{\beta})>e^{\bm{l}^{(j)}}_0(\tbt+t\bm{\beta})>e^{\bm{l}^{(j)}}_0(\tbt)=\max F^{k_0+1}_0$,
\item for the other $r-J_{\bm{\beta}}$ many $j$'s, we have $E^{k_0+r-s}(\tbt+t\bm{\beta})<e^{\bm{l}^{(j)}}_0(\tbt+t\bm{\beta})<e^{\bm{l}^{(j)}}_0(\tbt)=\min F^{k_0}_0$.
\end{itemize}
Thus 
\begin{align}\label{relation21}
J_{\bm{\beta}}=s+1.
\end{align}

On the other hand, we have that for $0>t>-\epsilon$ small enough,
\begin{itemize}
\item there are $r-J_{\bm{\beta}}-J_0$ many $j$'s such that $E^{k_0-s-1}(\tbt+t\bm{\beta})>e^{\bm{l}^{(j)}}_0(\tbt+t\bm{\beta})>e^{\bm{l}^{(j)}}_0(\tbt)=\max F^{k_0+1}_0$,
\item for the other $J_{\bm{\beta}}+J_0$ many $j$'s, we have $E^{k_0+r-s}(\tbt+t\bm{\beta})<e^{\bm{l}^{(j)}}_0(\tbt+t\bm{\beta})<e^{\bm{l}^{(j)}}_0(\tbt)=\min F^{k_0}_0$.
\end{itemize}
Thus 
\begin{align}\label{relation22}
J_{\bm{\beta}}+J_0=r-s-1.
\end{align}
Combining this with (\ref{relation21}), we have,
\begin{align}\label{relation23}
r-2s=J_0+2.
\end{align}
However, this contradicts with (\ref{relation13}), since $J_{\tbb}^0\geq 1$. $\hfill{} \Box$

\section{Proof of Lemma \ref{E=0}}\label{Eequal0sec}
The spirit of this proof is similar to that of Lemma \ref{Enot0}, but requires different choices of $\tbt, \bm{l}^{(1)}$ and $\bm{\beta}, \tbb$.

Without loss of generality, we assume $q_1$ is odd. 
We assume $q_i$'s, $i\geq 2$, are even, since otherwise, we could simply replace $q_i$ with $2q_i$, $i\geq 2$. 
Throughout this section, we will consider the case when $\min F^{k_0}_0=\max F^{k_0+1}=0$.

\subsection{$d=2$}\

This result has already been proved in \cite{EF}. Here we give an alternative self-contained proof. 

We let $\tbt=(\frac{1}{2q_1}, 0)$, $\bm{l}^{(1)}=(\frac{q_1-1}{2}, 0)$, and observe that 
\begin{align}
\left\lbrace
\begin{matrix}
0=2\cos{\pi}+2\cos{0}=e^{\bm{l}^{(1)}}_0(\tbt),\\
\bm{0}=\nabla e^{\bm{l}^{(1)}}_0(\tbt).
\end{matrix}
\right.
\end{align}

Again, we let $\bm{l}^{(2)}, ..., \bm{l}^{(r)}\in \Lambda$ (if any) to be {\it all} the vectors in $\Lambda$ such that $e^{\bm{l}^{(1)}}_0(\tbt)=e^{\bm{l}^{(2)}}_0(\tbt)=\cdots=e^{\bm{l}^{(r)}}_0(\tbt)=0$. 
Let $0\leq s\leq r-1$ be such that 
$E^{k_0-s-1}_0(\bthe)>E^{k_0-s}_0(\bthe)=\cdots=E^{k_0}_0(\bthe)=\cdots=E^{k_0+r-s-1}_0(\bthe)>E^{k_0+r-s}_0(\bthe)$ for any $\|\bthe-\tbt\|_{\Theta}<\epsilon$.

Let $\bm{l}^{(j)}$, $1\leq j\leq r$, be such that $\nabla e^{\bm{l}^{(j)}}_0(\tbt)=\bm{0}$. 
Then 
$\sin{2\pi(\tthe_1+\frac{l^{(j)}_1}{q_1})}=\sin{2\pi(\tthe_2+\frac{l^{(j)}_2}{q_2})}=0$. 
Taking into account that $e^{\bm{l}^{(j)}}_0(\tbt)=0$,
this implies $j=1$.
Hence the number of $j$'s such that $\nabla e^{\bm{l}^{(j)}}_0(\tbt)=\bm{0}$ is equal to $1$.

Now let $\bm{\beta}^+=(1, 0)$ and $\bm{\beta}^-=(0, 1)$. 
Let $J_{\bm{\beta}^\pm}$, $J^0_{\bm{\beta}^{\pm}}$ be as in the proof of Lemma \ref{Enot0}.

First, it is easy to see that $J_{\bm{\beta}^+}^0=J_{\bm{\beta}^-}^0=0$. Indeed, if there is $j$ such that $\nabla e^{\bm{l}^{(j)}}_0(\tbt)\neq \bm{0}$ and $\bm{\beta}^+\cdot \nabla e^{\bm{l}^{(j)}}_0(\tbt)=0$, then $\sin{2\pi (\tthe_1+\frac{l^{(j)}_1}{q_1})}=0$, which implies $\cos{2\pi (\tthe_1+\frac{l^{(j)}_1}{q_1})}=\pm 1$. This in turn implies $\cos{2\pi (\tthe_2+\frac{l^{(j)}_2}{q_2})}=\mp 1$, and hence $\nabla e^{\bm{l}^{(j)}}_0(\tbt)= \bm{0}$, contradiction. 
The case $J^0_{\bm{\beta}^-}=0$ can be argued in the same way.

Secondly, by (\ref{taylor21}), we have that for $|t|<\epsilon$ small enough,
\begin{align}\label{E0d2taylor2}
e^{\bm{l}^{(1)}}_0(\tbt+t\bm{\beta}^{\pm})
=&\pm 4\pi^2 t^2+O(t^3),
\end{align}
so $e^{\bm{l}^{(1)}}_0$ increases in the direction of $\bm{\beta}^+$ and decreases in the direction of $\bm{\beta}^-$.

Combining (\ref{taylort}) with (\ref{E0d2taylor2}) for $\bm{\beta}^+$, on one hand, we have, for $\epsilon>t>0$ small enough,
\begin{itemize}
\item there are $J_{\bm{\beta}^+}+1$ many $j$'s such that $E^{k_0-s-1}(\tbt+t\bm{\beta}^+)>e^{\bm{l}^{(j)}}_0(\tbt+t\bm{\beta}^+)>0=\max F^{k_0+1}_0$,
\item for the other $r-J_{\bm{\beta}^+}-1$ many $j$'s, we have $E^{k_0+r-s}(\tbt+t\bm{\beta}^+)<e^{\bm{l}^{(j)}}_0(\tbt+t\bm{\beta}^+)<0=\min F^{k_0}_0$.
\end{itemize}
Hence 
\begin{align}\label{E0d2relation1}
J_{\bm{\beta}^+}+1=s+1.
\end{align}
On the other hand, for $0>t>-\epsilon$ small enough, we have,
\begin{itemize}
\item there are $r-J_{\bm{\beta}^+}$ many $j$'s such that $E^{k_0-s-1}(\tbt+t\bm{\beta}^+)>e^{\bm{l}^{(j)}}_0(\tbt+t\bm{\beta}^+)>0=\max F^{k_0+1}_0$,
\item for the other $J_{\bm{\beta}^+}$ many $j$'s, we have $E^{k_0+r-s}(\tbt+t\bm{\beta}^+)<e^{\bm{l}^{(j)}}_0(\tbt+t\bm{\beta}^+)<0=\min F^{k_0}_0$.
\end{itemize}
Hence 
\begin{align}\label{E0d2relation2}
J_{\bm{\beta}^+}=r-s-1.
\end{align}
Thus combining (\ref{E0d2relation1}) with (\ref{E0d2relation2}), we have
\begin{align}\label{E0d2relation3}
r=2s+1.
\end{align}

Similarly, combining (\ref{taylort}) with (\ref{E0d2taylor2}) for $\bm{\beta}^-$, on one hand, we have, for $\epsilon>t>0$ small enough,
\begin{itemize}
\item there are $J_{\bm{\beta}^-}$ many $j$'s such that $E^{k_0-s-1}(\tbt+t\bm{\beta}^-)>e^{\bm{l}^{(j)}}_0(\tbt+t\bm{\beta}^-)>0=\max F^{k_0+1}_0$,
\item for the other $r-J_{\bm{\beta}^-}$ many $j$'s, we have $E^{k_0+r-s}(\tbt+t\bm{\beta}^-)<e^{\bm{l}^{(j)}}_0(\tbt+t\bm{\beta}^-)<0=\min F^{k_0}_0$.
\end{itemize}
Hence 
\begin{align}\label{E0d2relation4}
J_{\bm{\beta}^-}=s+1.
\end{align}
On the other hand, for $0>t>-\epsilon$ small enough, we have,
\begin{itemize}
\item there are $r-J_{\bm{\beta}^-}-1$ many $j$'s such that $E^{k_0-s-1}(\tbt+t\bm{\beta}^-)>e^{\bm{l}^{(j)}}_0(\tbt+t\bm{\beta}^-)>0=\max F^{k_0+1}_0$,
\item for the other $J_{\bm{\beta}^-}+1$ many $j$'s, we have $E^{k_0+r-s}(\tbt+t\bm{\beta}^-)<e^{\bm{l}^{(j)}}_0(\tbt+t\bm{\beta}^-)<0=\min F^{k_0}_0$.
\end{itemize}
Hence 
\begin{align}\label{E0d2relation5}
J_{\bm{\beta}^-}+1=r-s-1.
\end{align}
Thus combining (\ref{E0d2relation4}) with (\ref{E0d2relation5}), we have
\begin{align}\label{E0d2relation6}
r=2s+3.
\end{align}
This contradicts with (\ref{E0d2relation3}). $\hfill{} \Box$

\subsection{$d\geq 3$}\

Let us choose $\tbt, \bm{l}^{(1)}$ with $\tthe_1=\frac{1}{2q_1}, l^{(1)}_1=\frac{q_1-1}{2}$ and $\tthe_i, l^{(1)}_i$, $2\leq i\leq d$, be such that 
$\cos{2\pi(\tthe_i+\frac{l^{(1)}_i}{q_i})}=\frac{1}{d-1}<1$ and $\sin{2\pi (\tthe_i+\frac{l^{(1)}_i}{q_i})}>0$. 
Let $\bm{\beta}=(1, 0, 0,..., 0)$, then clearly we have,
\begin{align}\label{E0d3e1ortho}
\nabla e^{\bm{l}^{(1)}}_0(\tbt)\neq \bm{0}\ \ \mathrm{and}\ \ \bm{\beta}\cdot \nabla e^{\bm{l}^{(1)}}_0(\tbt)=0.
\end{align}

Let $\bm{l}^{(2)}, ..., \bm{l}^{(r)}\in \Lambda$ (if any) be {\it all} the vectors in $\Lambda$ such that $e^{\bm{l}^{(1)}}_0(\tbt)=e^{\bm{l}^{(2)}}_0(\tbt)=\cdots=e^{\bm{l}^{(r)}}_0(\tbt)$. 
Let $0\leq s\leq r-1$ be such that 
$E^{k_0-s-1}_0(\bthe)>E^{k_0-s}_0(\bthe)=\cdots=E^{k_0}_0(\bthe)=\cdots=E^{k_0+r-s-1}_0(\bthe)>E^{k_0+r-s}_0(\bthe)$ for any $\|\bthe-\tbt\|_{\Theta}<\epsilon$.

Let $J_0$, $J_{\bm{\beta}}$, $J_{\bm{\beta}}^0$ be as in the proof of Lemma \ref{Enot0}.
Then by (\ref{E0d3e1ortho}), $J_{\bm{\beta}}^0\geq 1$.

Clearly, for $J_0+J_{\bm{\beta}}^0$ many $j$'s, we have $\bm{\beta}\cdot \nabla e^{\bm{l}^{(j)}}_0(\tbt)=0$, which means $\sin{2\pi (\tthe_1+\frac{l^{(j)}_1}{q_1})}=0$. 
Since our $\tthe_1$ equals $\frac{1}{2q_1}$, we must have 
\begin{align}\label{E0d3first=-2}
\cos{2\pi (\tthe_1+\frac{l^{(j)}_1}{q_1})}=-1.
\end{align} 


Thus, by (\ref{taylor21}) and (\ref{E0d3first=-2}), we have that for $j$ (in total $J_0+J_{\bm{\beta}}^0$ many) such that $\bm{\beta}\cdot \nabla e^{\bm{l}^{(j)}}_0(\tbt)=0$, for $|t|<\epsilon$ small enough,
\begin{align}\label{E0d3taylor2}
e^{\bm{l}^{(j)}}_0(\tbt+t\bm{\beta})
=&e^{\bm{l}^{(j)}}_0(\tbt)+\frac{t^2}{2} \left(-8\pi^2 \cos{2\pi (\tthe_1+\frac{l^{(j)}_1}{q_1})} \right)+O(t^3)\notag \\
=&4\pi^2 t^2+O(t^3) \notag\\
>&0.
\end{align}
Hence, combining (\ref{taylort}) with (\ref{E0d3taylor2}), on one hand, we have, for $\epsilon>t>0$ small enough,
\begin{itemize}
\item there are $J_{\bm{\beta}}+J_0+J_{\bm{\beta}}^0$ many $j$'s such that $E^{k_0-s-1}(\tbt+t\bm{\beta})>e^{\bm{l}^{(j)}}_0(\tbt+t\bm{\beta})>0=\max F^{k_0+1}_0$,
\item for the other $r-J_{\bm{\beta}}-J_0-J_{\bm{\beta}}^0$ many $j$'s, we have $E^{k_0+r-s}(\tbt+t\bm{\beta})<e^{\bm{l}^{(j)}}_0(\tbt+t\bm{\beta})<0=\min F^{k_0}_0$.
\end{itemize}
Hence 
\begin{align}\label{E0d3relation1}
J_{\bm{\beta}}+J_0+J_{\bm{\beta}}^0=s+1.
\end{align}
On the other hand, for $0>t>-\epsilon$ small enough, we have,
\begin{itemize}
\item there are $r-J_{\bm{\beta}}$ many $j$'s such that $E^{k_0-s-1}(\tbt+t\bm{\beta})>e^{\bm{l}^{(j)}}_0(\tbt+t\bm{\beta})>0=\max F^{k_0+1}_0$,
\item for the other $J_{\bm{\beta}}$ many $j$'s, we have $E^{k_0+r-s}(\tbt+t\bm{\beta})<e^{\bm{l}^{(j)}}_0(\tbt+t\bm{\beta})<0=\min F^{k_0}_0$.
\end{itemize}
Hence 
\begin{align}\label{E0d3relation2}
J_{\bm{\beta}}=r-s-1.
\end{align}
Thus combining (\ref{E0d2relation1}) with (\ref{E0d2relation2}), we have
\begin{align}\label{E0d3relation3}
2s-r=J_0+J_{\bm{\beta}}^0-2.
\end{align}

Now we choose $\tbb\in \R^d$, $\|\tbb\|_{\R^d}=1$, such that $\tbb\cdot \nabla e^{\bm{l}^{(j)}}_0(\tbt)\neq 0$ for any $1\leq j\leq r$ with 
$\nabla e^{\bm{l}^{(j)}}_0(\tbt)\neq \bm{0}$, and satisfies the following:
\begin{align}\label{E0tbb-bb}
1-\tilde{\beta}_1^2+\sum_{i=2}^{d} \tilde{\beta}_i^2<\frac{1}{2}.
\end{align}
This inequality essentially says $\tbb\sim \bm{\beta}$.

With $J_{\tbb}$ defined as before, by (\ref{taylor21}), (\ref{E0d3first=-2}) and (\ref{E0tbb-bb}), we have that for $j$ (in total $J_0$ many) such that $\nabla e^{\bm{l}^{(j)}}_0(\tbt)=\bm{0}$, for $|t|<\epsilon$ small enough,
\begin{align}
e^{\bm{l}^{(j)}}_0(\tbt+t\tbb)
=&\frac{t^2}{2} \left(8\pi^2-8\pi^2(1-\tilde{\beta}_1^2)-8\pi^2 \sum_{i=2}^d \cos{2\pi (\tthe_i+\frac{l^{(j)}_i}{q_i})} \tilde{\beta}_i^2 \right) +O(t^3) \notag\\
>&{2\pi^2} t^2 +O(t^3)>0.\label{E0d3taylor4}
\end{align}

As before, combining (\ref{taylort}) with (\ref{E0d3taylor4}), on one hand, we have, for $\epsilon>t>0$ small enough,
\begin{itemize}
\item there are $J_0+J_{\tbb}$ many $j$'s such that $E^{k_0-s-1}(\tbt+t\tbb)>e^{\bm{l}^{(j)}}_0(\tbt+t\tbb)>0=\max F^{k_0+1}_0$,
\item for the other $r-J_0-J_{\tbb}$ many $j$'s, we have $E^{k_0+r-s}(\tbt+t\tbb)<e^{\bm{l}^{(j)}}_0(\tbt+t\tbb)<0=\min F^{k_0}_0$.
\end{itemize}
Hence 
\begin{align}\label{E0d3relation4}
J_0+J_{\tbb}=s+1.
\end{align}
On the other hand, for $0>t>-\epsilon$ small enough, we have,
\begin{itemize}
\item there are $r-J_{\tbb}$ many $j$'s such that $E^{k_0-s-1}(\tbt+t\tbb)>e^{\bm{l}^{(j)}}_0(\tbt+t\tbb)>0=\max F^{k_0+1}_0$,
\item for the other $J_{\tbb}$ many $j$'s, we have $E^{k_0+r-s}(\tbt+t\tbb)<e^{\bm{l}^{(j)}}_0(\tbt+t\tbb)<0=\min F^{k_0}_0$.
\end{itemize}
Hence 
\begin{align}\label{E0d3relation5}
J_{\tbb}=r-s-1.
\end{align}
Thus combining (\ref{E0d3relation4}) with (\ref{E0d3relation5}), we have
\begin{align}\label{E0d3relation6}
2s-r=J_0-2.
\end{align}
This contradicts (\ref{E0d3relation3}) since $J_{\bm{\beta}}^0\geq 1$. $\hfill{} \Box$

\section{Example with exactly two intervals}\label{countersec}
Let all the $q_i$'s be even and $\delta>0$ be any small positive number. 
We are going to construct $V$ with minimal period $\bm{q}$, 
such that $\|V\|_{\infty}=\delta$ and the spectrum of $H_{V}$ does not contain the point $0$. 
This example is a modification of Kr\"uger's example (see Theorem 6.3 in \cite{Kru}), where $V$ is $(2,2,...,2)$-periodic.

Let us define 
\begin{align}
V_{\bm{q}}(\bm{n})=
\left\lbrace
\begin{matrix}
(1-{\delta^2}/{d})\delta\ \ \ \ \ \ \ \ &\mathrm{if}\ \bm{n}\equiv \bm{0}\ \mathrm{(}\mod\ \bm{q}\mathrm{)}\\
&\\
\delta (-1)^{|\bm{n}|}\ \ \ \  \ &\mathrm{otherwise}
\end{matrix}
\right.
\end{align}
It can be easily checked that $V_{\bm{q}}$ has minimal period $\bm{q}$ and $\|V\|_{\infty}=\delta$. 
The fact that the spectrum of $H_{V}$ does not contain $0$ will follow from the following lemma.
\begin{lem}\label{counterlemma}
There exists constant $\delta_0>0$ such that for any $0<\delta<\delta_0$, we have
\begin{align*}
\|(H_0+ V_{\bm{q}})u\|> \frac{1}{2}\delta
\end{align*}
holds for any unit vector $u\in l^2(\Z^d)$.
\end{lem}

\subsection*{Proof of Lemma \ref{counterlemma}}
Let us consider
\begin{align}\label{c0}
\|(H_0+ V_{\bm{q}})u\|^2=\|H_0 u\|^2+\|V_{\bm{q}} u\|^2+2 (H_0 u, V_{\bm{q}}u)\geq \|V_{\bm{q}}u\|^2+2 (H_0 u, V_{\bm{q}}u),
\end{align}
in which the first term obviously satisfies 
\begin{align}\label{c1}
\|V_{\bm{q}}u\|^2=\sum_{\bm{n}\in \Z^d} |V_{\bm{q}}(\bm{n})|^2 |u(\bm{n})|^2\geq (1-\delta^2/d)^2 \delta^2\geq (1-\delta^2)^2 \delta^2.
\end{align}
Let $\{\bm{b}_i\}$ be the standard basis for $\R^d$.
The second term in (\ref{c0}) could be estimated in the following way:
\begin{align}\label{c2}
(H_0 u, V_{\bm{q}}u)
=&\sum_{\bm{n}\in \Z^d}\left(\sum_{i=1}^d u(\bm{n}\pm \bm{b}_i)\right)V_{\bm{q}}(\bm{n}) u({\bm{n}}) \notag\\
=&\sum_{i=1}^d \sum_{\bm{n}\in \Z^d} u(\bm{n}+\bm{b}_i)u(\bm{n}) (V_{\bm{q}}(\bm{n})+V_{\bm{q}}(\bm{n}+\bm{b}_i)).
\end{align}
Note that by our construction and the fact that $q_i$'s are even, 
\begin{align}\label{c3}
V_{\bm{q}}(\bm{n})+V_{\bm{q}}(\bm{n}+\bm{b}_i)=
\left\lbrace
\begin{matrix}
-\delta^3/d\ \ &\mathrm{if}\ \bm{n}\equiv -\bm{b}_i\ \mathrm{or}\ \bm{0}\ \mathrm{(mod}\ \bm{q}\mathrm{)}\\
0 &\mathrm{otherwise}
\end{matrix}
\right.
\end{align}
Combining (\ref{c2}) with (\ref{c3}), we get
\begin{align}\label{c4}
|(H_0 u, V_{\bm{q}} u)|\leq  \frac{\delta^3}{d} \sum_{i=1}^d \sum_{\bm{n}\in \Z^d} |u(\bm{n}+\bm{b}_i)| |u(\bm{n})|\leq \delta^3.
\end{align}
Now combining (\ref{c0}), (\ref{c1}) with (\ref{c4}), we get
\begin{align*}
\|(H_0+ V_{\bm{q}})u\|^2\geq (1-\delta^2)^2 \delta^2-2 \delta^3>\frac{1}{4}\delta^2,
\end{align*}
provided $\delta$ small.
$\hfill{} \Box$

\appendix
\section{\\}
\subsection*{Proof of Lemma \ref{thereexiststheta}}
Without loss of generality we could assume $E\geq 0$.

If $d=2\tilde{d}$ is an even number, then we could take $(0, 1/2)\ni \theta_1=\cdots =\theta_{\tilde{d}}=1-\theta_{\tilde{d}+1}=\cdots=1-\theta_{2\tilde{d}}$ be such that 
$\cos{2\pi \theta_1}=\frac{E}{4\tilde{d}}\neq \pm1$.

If $d=2\tilde{d}+1$ is an odd number and $E\in [2, 4\tilde{d}+2)$, then we could take $\theta_{2\tilde{d}+1}=0$ and \linebreak $(0, 1/2)\ni \theta_1=\cdots =\theta_{\tilde{d}}=1-\theta_{\tilde{d}+1}=\cdots=1-\theta_{2\tilde{d}}$ be such that 
$\cos{2\pi \theta_1}=\frac{E-2}{4\tilde{d}}\neq \pm 1$.

If $d=2\tilde{d}+1$ is an odd number and $E\in [0, 2)$, then we could take $\theta_{2\tilde{d}+1}=\frac{1}{2}$ and \linebreak $(0, 1/2)\ni \theta_1=\cdots =\theta_{\tilde{d}}=1-\theta_{\tilde{d}+1}=\cdots=1-\theta_{2\tilde{d}}$ be such that 
$\cos{2\pi \theta_1}=\frac{E+2}{4\tilde{d}}\neq \pm 1$. $\hfill{} \Box$

\section*{Acknowledgement}
This research was partially supported by the NSF DMS–1401204.

\bibliographystyle{amsplain}

\begin{thebibliography}{10}
\bibitem{DT}B. E. J. Dahlberg and E. Trubowitz, A remark on two dimensional periodic potentials,
Comment. Math. Helvetici 57 (1982), 130–134.


\bibitem{EF}M. Embree and J. Fillman,  
Spectra of Discrete Two-Dimensional Periodic Schr\"odinger Operators with Small Potentials.
arXiv:1701.00863

\bibitem{HM}B. Helffer and A. Mohamed, Asymptotics of the density of states for the Schr\"odinger
operator with periodic electric potential, Duke Math. J. 92 (1998), 1–60.

\bibitem{Kar}Y. E. Karpeshina, Perturbation Theory for the Schr\"odinger Operator with a Periodic
Potential, Lecture Notes in Math. Vol. 1663, Springer Berlin 1997.

\bibitem{Kru}H. Kr\"uger, Periodic and limit-periodic discrete Schr\"odinger operators. arXiv:1108.1584.


\bibitem{P}L. Parnovski, Bethe-Sommerfeld conjecture, Ann. Henri Poincar\'{e} 9 (2008), 457–508.


\bibitem{PSDuke}L. Parnovski and A. V. Sobolev, Bethe-Sommerfeld Conjecture for Polyharmonic
Operators, Duke Math. J., 2001.

\bibitem{PSAHP}L. Parnovski and A. V. Sobolev, Perturbation Theory and the Bethe-Sommerfeld
Conjecture, Annals H. Poincar\'{e}, 2001.


\bibitem{PS}V. N. Popov and M. Skriganov, A remark on the spectral structure of the two dimensional
Schr\"odinger operator with a periodic potential, Zap. Nauchn. Sem. LOMI AN
SSSR 109 (1981), 131–133 (Russian).

\bibitem{S79}M. Skriganov, Proof of the Bethe-Sommerfeld conjecture in dimension two, Soviet
Math. Dokl. 20 (1979), 1, 89–90.

\bibitem{S84}M. Skriganov, Geometrical and Arithmetical Methods in the Spectral Theory of the
Multi-dimensional Periodic Operators, Proc. Steklov Math. Inst. Vol. 171, 1984.

\bibitem{S85} M. Skriganov, The spectrum band structure of the three-dimensional Schr\"odinger
operator with periodic potential, Inv. Math. 80 (1985), 107–121.

\bibitem{V}O. A. Veliev, Perturbation theory for the periodic multidimensional Schr\"odinger operator
and the Bethe-Sommerfeld Conjecture, Int. J. Contemp. Math. Sci. 2 (2007),
no. 2, 19–87.

\end{thebibliography}

\end{document}